\documentclass[twocolumn,showpacs,amsmath,amssymb]{revtex4}

\usepackage{graphicx}
\usepackage{dcolumn}
\usepackage{bm}
\usepackage[dvipdfm]{xcolor}

\begin{document}



\title{Density Operator Description of Atomic Ordered Spatial Modes in Cavity QED}

\author{Zhen Fang}
\author{Baoguo Yang}
\author{Xuzong Chen}
\author{Xiaoji Zhou}
\email[E-mail: ]{xjzhou@pku.edu.cn}
\address{School of Electronics Engineering $\&$ Computer Science, Peking University, Beijing 100871,China}


\begin{abstract}
We present a quantum Monte-Carlo simulation for a pumped atom in a strong coupling cavity with dissipation, where two ordered spatial modes are formed for the atomic probability density, with the peaks distributed either only in the odd sites or only in the even ones of the lattice formed by the cavity field. A mixed state density operator model, which describes the coupling between different atomic spatial modes and the corresponding cavity field components, is proposed, which goes beyond the pure state interpretation. We develop a new decomposition treatment to derive the atomic spatial modes as well as the cavity field statistics from the simulation results for the steady state. With this treatment, we also investigate the dynamical process for the probabilities of the atomic spatial modes in the adiabatic limit. According to the analysis of the fitting error between the simulation results and the density operator model, the latter is a good description for the system.
\end{abstract}

.


\maketitle

\section{Introduction}
Laser light has been widely used for manipulating cold or ultracold atomic samples~\cite{rmp70}. Such experiments include the quantum phase transition from superfluid phase to Mott insulator phase in optical lattices~\cite{nature415}, free-space or cavity-enhanced superradiant scattering of Bose-Einstein condensates (BEC)~\cite{science285, prl98p1, pra78}, and collective atomic recoil lasing (CARL)~\cite{prl98p1, nim341, pra80}. During the last decade, strong light-matter coupling regime in cavity quantum electrodynamics (cavity QED) becomes accessible, making the cavity QED system an important platform for studying the interaction between a quantized light field and atomic samples~\cite{fsqo}. In such a system the photons make several round trips before they decay out of the cavity, and the back-action of the atomic distribution on the light field is important. Because of the fast response and high sensitivity to optical fields, cavity QED systems have been used as sensors at the quantum level in many experiments, such as quantum nondemolition measurement~\cite{prl102}, atomic dynamics detection~\cite{ol21}, atomic quantum phase probing~\cite{prl98p2} and quantum optomechanics~\cite{nat452, prl103}. Some new phenomena, such as cavity cooling~\cite{prl79, jpb38}, atomic self-organization~\cite{prl89, prl91, oc273, epjd48}, and Dicke model phase transition~\cite{nature464, sci336, prl104} also appear in cavity QED systems.

When an atom is pumped by a far-detuned standing-wave laser perpendicular to the cavity axis,two ordered spatial modes are quickly formed, with the peaks of the probability density distributed either only in the odd sites or only in the even ones. These two modes scatter photons with opposite phases, and the scattered photons form an optical lattice potential in the cavity. With quantum fluctuations and cavity field amplitude collapses the atomic spatial distribution may reordered into one of the two modes or back into this symmetric composite mode, which is known well as the atomic self-organization~\cite{prl89, prl91, oc273, epjd48, epjd46}. Although in most researches the decay through the cavity mirror is considered, the state for the atom-cavity system is still interpreted by a pure entangled state given as $|{\rm odd} \rangle |\alpha \rangle + |{\rm even} \rangle |-\alpha \rangle$, where $|{\rm odd} \rangle$ and $|{\rm even}\rangle$ are states of the atomic ordered modes, and $|\alpha\rangle$ and $|-\alpha\rangle$ are states of the cavity field~\cite{oc273, epjd46}. Indeed, for a short time at the beginning, the effect of the cavity decay is small, and the state of the system can be regarded as a pure state. However, this transient state is far different from the steady state of the system. After the evolution over a long time, the dissipation process plays an important role~\cite{pra82}, and the system may no longer be a pure state. Therefore, it is more reasonable by describing the system with a mixed state density operator rather than a pure state. Moreover, as mentioned in Ref.~\cite{epjd46}, there should be a vacuum state component for the steady state, and an atomic spatial mode, which we call ``the residual mode'', is together with this state component. Besides, the quantitative study on the proportions of the atomic modes is still missing. Furthermore, the time evolution for the weights for these three modes, that is, the dynamical process for the growth of the ordered modes and the decrease of the residual mode, has not been given quantitatively, with the presence of the cavity decay. In order to clarify these problems, the model we investigated in this paper is slightly different from the ones by other researchers~\cite{oc273, epjd46, prl95, njp9}. Here the atomic sample is confined by, for example, an external magnetic trap, instead of a strong optical lattice. The average photon number in the cavity can be very low, and the quantum fluctuations of the cavity field are notable. The tunneling of the atom between neighboring lattice sites can be significant, and the atomic spatial distribution is not so well localized as has been investigated in many researches. In this situation it is not valid to consider only the lowest vibrational energy band in the Wannier expansion of the atomic wave function. Therefore, a Bose-Hubbard-type Hamiltonian can no longer be implemented, and a fully quantum mechanical model is required to describe such a system. The steady state and dynamical properties are described by corresponding density operators. Instead of investigating the dynamics for atomic self-organization (or the Dicke model phase transition), this model is appropriate for studying the steady state properties and the time evolution of the atomic spatial modes, which is the main purpose of this paper.

This paper is organized as follows. In Section 2, we present a fully quantum mechanical model in first quantization form, which describes a coherently-pumped atom (or a non-interacting BEC) strongly coupled to a weak cavity field with dissipation through the cavity mirrors. In Section 3, we first introduce the Monte-Carlo wave function method, which is used to simulate the stochastic evolution of the state vector of the system~\cite{pra82, prl68, epjd44}, and from which we can derive the stationary and time-dependent density operators to describe the steady state and time evolution properties of the system. By analyzing the atomic coherence and the cavity field properties of the simulation results, we give a mixed state density operator model describing the coupling between atomic spatial modes and corresponding cavity field components. Based on this model, we develop a mathematical treatment to decompose the atomic modes from the simulated density operator. According to the analysis of the fitting accuracy, this model density operator is a good fitting to the simulated results. In Section 4, by acting this decomposition treatment on the time-dependent density operator, we investigate the time evolution of the atomic spatial modes, from which we show how the atomic ordered modes are formed with the establishment of the cavity field. The conclusion is given in Section 5.

\section{Model of a transversally-pumped atom in the cavity}

\begin{figure}[tbp]
\begin{center}
\includegraphics[width=6cm]{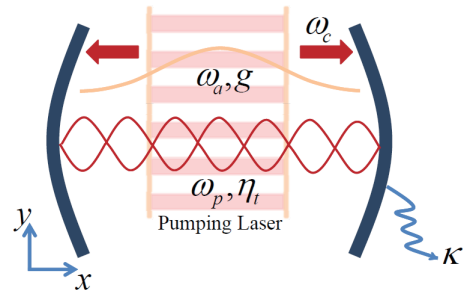}
\end{center}
\caption{(Color online) The atom pump scheme. A two-level atom with transition frequency $\omega_{\rm a}$ is illuminated by a standing-wave laser with frequency $\omega_{\rm p}$ directly in a strong coupling cavity with resonance frequency $\omega_{\rm c}$. The photons can be scattered into the cavity by the atom and an optical lattice can be formed. The strength of the pumping laser is $\eta_{\rm t}$. The cavity decay rate is $2\kappa$ and the coupling strength between the atom and the cavity field is $g$. For simplicity, in our calculation the cavity is approximated as a 1D box for confining the cavity photons along its axial direction.}
\label{fig:atom_pump_scheme}
\end{figure}

We start with a two-level atom with mass $\mu$ and transition frequency $\omega_{\rm a}$ strongly coupled to a single-mode cavity field with frequency $\omega_{\rm c}$. The upper and lower states of the atom are represented as $\left| {\rm e} \right\rangle$ and $\left| {\rm g} \right\rangle$, respectively. The coupling strength between the atom and the cavity field is $g$. A plane standing-wave laser with frequency $\omega_{\rm p}$ and pumping strength $\eta_{\rm t}$, which is perpendicular to the cavity axis, illuminates the atom directly. This scheme is often referred to as the atomic or transverse pumping (see Fig.~\ref{fig:atom_pump_scheme}). The wave numbers are $K$ for the cavity field and $k_{\rm p}$ for the pumping field, and the atomic recoil frequency by scattering a cavity photon is $\omega_{\rm r}=\hbar K^2 / (2 \mu)$. Using the rotating-wave and electric-dipole approximations, the Jaynes-Cummings Hamiltonian of the system can be depicted in the frame rotating with $\omega_{\rm p}$ as
\begin{eqnarray}\label{eq:jc_hamiltonian}
&H& = - \hbar \Delta_{{\rm c}} \hat a^\dag \hat a + \dfrac{\hat {\mathbf p}^2}{2 \mu} + V(\hat {\mathbf r}) - \hbar \Delta_{{\rm a}} \hat \sigma_+ \hat \sigma_-\nonumber\\
&&- i \hbar g f(\hat {\mathbf r}) \left( \hat \sigma_+ \hat a - \hat \sigma_- \hat a^\dag \right) - i \hbar \eta_{{\rm t}} \zeta(\hat {\mathbf r}) \left( \hat \sigma_+ - \hat \sigma_- \right),
\end{eqnarray}
where the terms on the right side describe free cavity field, atomic kinetic energy, external trapping potential, atomic excitation, coupling between the atom and the cavity field, and pumping of the atom by the transverse laser, respectively. $\Delta_{\rm c}=\omega_{\rm p}-\omega_{\rm c}$ and $\Delta_{\rm a}=\omega_{\rm p}-\omega_{\rm a}$ are the cavity and atomic detunings from the frequency of the pumping laser, respectively. $f(\hat {\mathbf r})$ is the cavity mode function and $\zeta(\hat {\mathbf r})$ is the pumping laser mode function. $\hat a$ and $\hat a^\dag$ are the annihilation and creation operators of the cavity field. $\hat \sigma_+ = | {\rm e} \rangle \langle {\rm g} |$ and $\hat \sigma_- = | {\rm g} \rangle \langle {\rm e} |$ are the raising and lowering operators of the atom. As it is difficult to treat a 2D model numerically due to high dimensions of the momentum space, we assume the potential $V(\hat {\mathbf r})$ is very tight in the $y-z$ plane and very loose along $x$ direction(in our work it is treated as a constant potential $V_0$). Therefore, the atomic motion is in fact restricted along the cavity axis ($x$ direction), and our model is reduced to 1D. Such an assumption has been discussed in researches about atomic self-organization~\cite{oc273, epjd46}. The cavity in our model is regarded as a 1D box, which confines the photons in the cavity along its axial direction. The cavity field is taken to be a sinusoidal plane wave $f(\hat {\mathbf r})=f(\hat x)=\sin \left( Kx \right)$. The mode function of the transverse laser is approximated by $\zeta(\hat {\mathbf r})=\cos \left( k_{\rm p}y \right)$. With 1D assumption (where the atomic motion is confined along $y=0$), the mode function of the transverse laser equals to 1.

In the case of far-off-resonance pumping, the large atomic detuning leads to weak atomic excitation, and the atomic spontaneous emission is negligible. However, the cavity loss with decay rate $2 \kappa$ still needs to be considered in this model, which is the dominant dissipation process for the system. From Eq.~(\ref{eq:jc_hamiltonian}) we can write the Heisenberg equations for $\hat \sigma_-$ and $\hat a$ with the dissipation included
\begin{eqnarray}\label{eq:heisenbery_sig}
\frac{{\rm d} \hat \sigma_-}{{\rm d} t} &=& i\Delta_{\rm a} \hat \sigma_- - (g f(\hat x)\hat a + \eta_{\rm t}),
\end{eqnarray}
\begin{eqnarray}\label{eq:heisenbery_a}
\frac{{\rm d} \hat a}{{\rm d} t} &=& \left(i\Delta_{\rm c}- \kappa\right) \hat a + g f(\hat x) \sigma_-.
\end{eqnarray}

Due to low saturation, we can adiabatically eliminate the atomic internal degrees of freedom, and the lowering operator of the atom is then presented as
\begin{eqnarray}
\hat \sigma_- \approx \frac{g f(\hat x) \hat a + \eta_{\rm t}}{i \Delta_{\rm a}},
\end{eqnarray}
with the raising operator $\hat \sigma_+=\hat \sigma_-^\dag$. We put these expressions back into Eq.~({\ref{eq:jc_hamiltonian}) to obtain the effective Hamiltonian for this 1D model
\begin{eqnarray}\label{eq:h_eff}
&H_{{\rm eff}}& = -\hbar \Delta_{\rm c} \hat a^\dag \hat a + \dfrac{\hat p^2}{2 \mu} + \hbar U_0 \sin^2 (K \hat x) \hat a^\dag \hat a\nonumber\\
&&+ \hbar U_{\rm t} \sin (K \hat x) \left( \hat a^\dag + \hat a \right).
\end{eqnarray}
In the effective Hamiltonian the terms describing the external trapping potential $V_0$ and the coupling between the atom and the pumping laser $\hbar \eta_{\rm t}^2/\Delta_a$, which are both constant terms, are neglected. $U_0=g^2/\Delta_{\rm a}$ is the effective coupling strength between the atom and the cavity field, and $U_{\rm t}=g \eta_{\rm t}/\Delta_{\rm a}$ is the effective pumping strength of the cavity field by atomic scattering.

Finally, we can write the master equation for the system
\begin{eqnarray}\label{eq:master_eq}
\dot \rho = \frac{1}{i \hbar} [H_{{\rm eff}}, \rho] + {\mathcal L} \rho.
\end{eqnarray}
Here $\rho$ is the density operator of the system. The Liouvillean term can be depicted as
\begin{eqnarray}
{\mathcal L} \rho = 2 \kappa \hat a \rho \hat a^\dag - \kappa [\hat a^\dag \hat a, \rho]_+  = \hat J_{\rm c} \rho \hat J_{\rm c}^\dag - \frac{1}{2}[\hat J_{\rm c}^\dag \hat J_{\rm c}, \rho]_+,
\end{eqnarray}
with the effect of cavity loss described by the jump operator $\hat J_{\rm c}=\sqrt{2 \kappa} \hat a$.

\section{Mixed state density operator of the atom-cavity system}

\subsection{Quantum Monte-Carlo wave function simulation}

In our model, the state of the atom is given in the basis $\{|k\rangle\}$, where $|k\rangle$ denotes the $k$th atomic momentum state with a momenta $p = \hbar k K$. The state of the cavity field is expressed in Fock basis $\{|n \rangle\}$, with $|n\rangle$ the number state of $n$ photons.  Since the minimal change of the atomic momentum is $\pm K$ by absorbing or emitting a cavity photon, it is physically reasonable to divide the momentum space with a step $K$. The state vector of the system is then given by $| \psi(t) \rangle = \sum_{k,n} C_{k,n}(t)|k,n\rangle$. Because of the atomic momentum diffusion in the periodic potential, we need a very large dimension for describing the atomic momentum space, which is taken to be $2^6$ from $-32\hbar K$ to $32 \hbar K$ in our simulation. The Fock basis of the cavity field state is truncated up to the 10th state, which is reasonable as long as the average photon number in the cavity is very small. Initially, the cavity field is in the vacuum state, and the atom is in the zero-momentum state. If we neglect all kinds of fluctuations that might exist in real experimental circumstance, the system has a spatial period $Kx=2\pi$, and the atomic initial state has translational symmetry. Therefore, the state vector of the system also has a period of $2\pi$, and we can consider a model which consists of only two lattice sites with periodic boundary conditions. This is the simplest system that can reveal the physics during the formation of the atomic ordered spatial modes.

In order to investigate the time evolution of the system presented in the previous section, we need to solve the master equation~(\ref{eq:master_eq}) and get the time dependence for each matrix element $\rho_{k,n;k^\prime,n^\prime}(t)$. Unfortunately, as the dimension of the Hibert space for the state vector is very large, it requires great effort for this numerical computation (in our simulation we need to solve $409600$ differential equations). For reducing the numerical difficulty, we use the Monte-Carlo wave function (MCWF) method to simulate the stochastic time evolution of the state vector $\left| \psi(t) \right\rangle$~\cite{prl68, epjd44}. During a time step $\delta t$, the probability for the cavity decay is $P_{\rm c} = 2\kappa \langle \hat a^\dag \hat a \rangle \delta t$, where $\delta t$ has to be chosen to satisfy $P_{\rm c} \ll 1$ to ensure the validity of the MCWF method. A random number $\epsilon$ between $0$ and $1$ is generated to justify whether a cavity decay event happens. If $\epsilon > P_{\rm c}$, no photon decays out of the cavity. The state vector evolves according to the Schr{\" o}dinger equation governed by the non-Hermitian Hamiltonian
\begin{eqnarray}
H_{\rm nH} = H_{\rm eff}-\frac{i}{2}\hat J_{\rm c}^\dag \hat J_{\rm c},
\end{eqnarray}\label{eq:h_nh}
and after a time step $\delta t$, the state vector has the form
\begin{eqnarray}
|\psi(t+\delta t)\rangle = \frac{1}{\sqrt{1-P_{\rm c}}}\left(1-i\frac{\delta t}{\hbar} H_{\rm nH}\right)|\psi(t)\rangle.
\end{eqnarray}
On the contrary, if $\epsilon < P_{\rm c}$, the state vector is
\begin{eqnarray}
|\psi(t+\delta t)\rangle = \frac{\sqrt{\delta t}}{\sqrt{P_{\rm c}}} \hat J_{\rm c}|\psi(t)\rangle.
\end{eqnarray}
It has been proved that the density operator $\rho(t)$ derived by averaging $| \psi(t) \rangle \langle \psi(t)|$ over several stochastic trajectories of $|\psi(t)\rangle$ evolves according to the standard master equation~\cite{prl68}.  According to the ergodic hypothesis, the steady-state property of the system can be expressed by running only one trajectory $\left| \psi (t) \right\rangle$ for a long time $T$ and calculating the steady-state density operator $\rho_{{\rm ss}}=(1/T) \int_{T_{{\rm rel}}}^T {\rm d}t \left| \psi (t) \right\rangle \left\langle \psi (t) \right|$~\cite{jpb38, pra82}. Here the influence of the initial transients is neglected, as $T$ is very large compared with $T_{\rm rel}$. In our simulation we choose $\kappa T = 25000$, which is much larger than $\kappa T_{\rm rel} \approx 20$. The dynamical process of the system can be expressed by running several ($N=200$ in our simulation) trajectories $\left| \psi_i (t) \right\rangle$ and calculating the time-dependent density operator $\rho(t) = (1/N) \sum_{i=1}^N \left| \psi_i (t) \right\rangle \left\langle \psi_i (t) \right|$.

\subsection{Model density operator describing the mixed state}

Based on the results of MCWF simulation, we will present a model density operator, which can giver a better interpretation for the steady state of the system compared to the pure state description. With this model, we are able to decompose the steady state density operator and get the density matrix for each atomic ordered mode with corresponding cavity field component.

\begin{figure}[tbp]
 \begin{center}
  \begin{picture}(0,0)
   \put(-10,80){(a)}\quad
  \end{picture}
  \includegraphics[width=3.9cm]{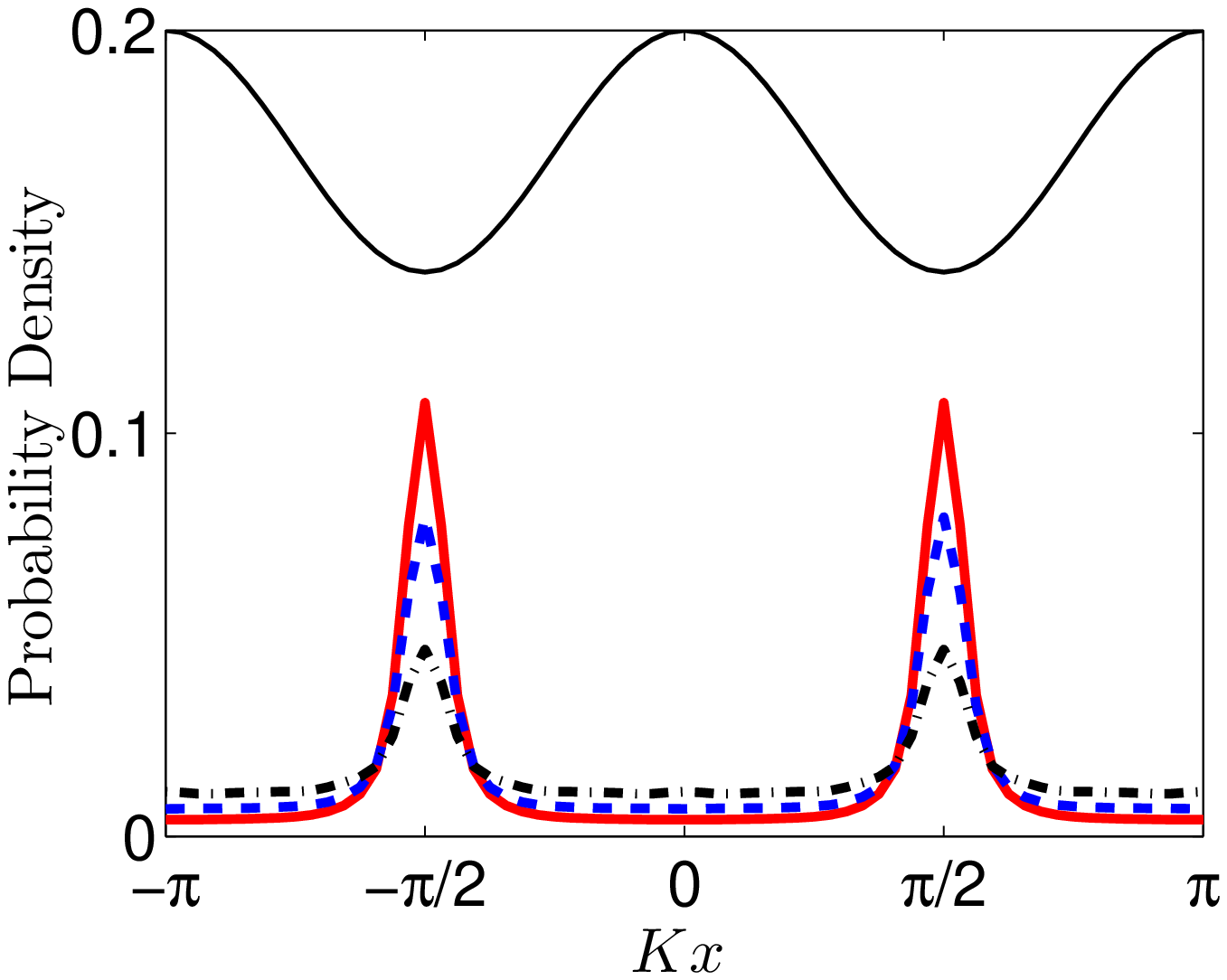}\quad\quad
  \begin{picture}(0,0)
   \put(-10,80){(b)}\quad
  \end{picture}
  \includegraphics[width=3.9cm]{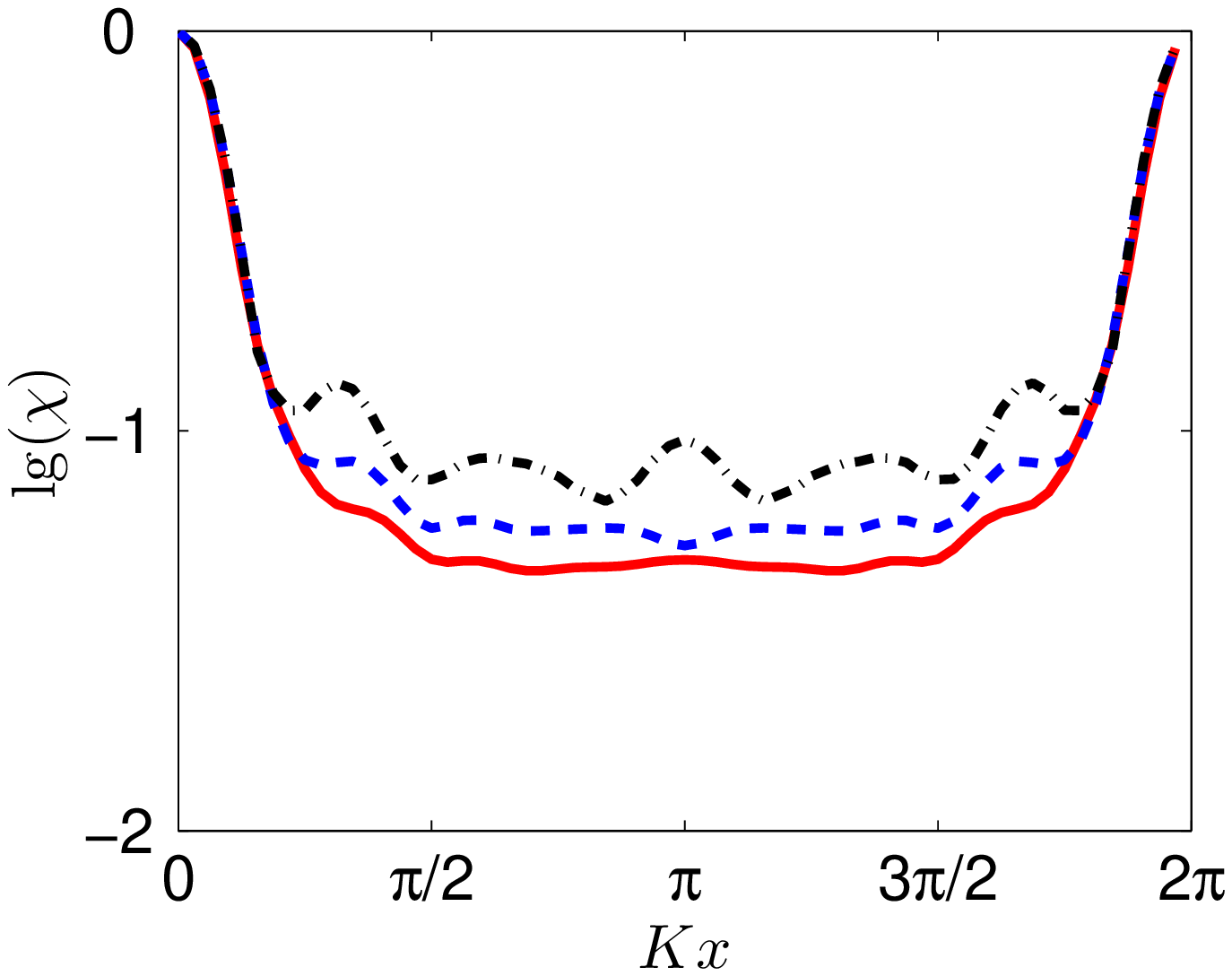}\quad\quad
 \end{center}
\caption{(Color online) (a) Atomic probability distribution in coordinate space for the steady state. The solid sinusoidal curve shows the spatial distribution of the lattice. (b) The common logarithm of the atomic spatial correlation function $\chi(x)$ for steady state. All results are calculated from the steady state density operator. The dash-dotted, dashed and solid curves correspond to $U_{\rm t}=-22\omega_{\rm r}$, $-38\omega_{\rm r}$ and $-49\omega_{\rm r}$, respectively. Other parameters are: $\kappa=31.25\omega_r$, $\Delta_c = U_0 = -390\omega_r$.} \label{fig:pn_coh_evol}
\end{figure}

The last term of the effective Hamiltonian (\ref{eq:h_eff}), which describes the scattering process, is the key point in the formation of the atomic ordered modes. The sinusoidal expression of the atomic spatial coordinate in this term indicates that the scattering process strongly depends on the atomic spatial distribution. For a symmetric atomic distribution in the odd and even sites, where the atom scatters photons with opposite phases, the resulting cavity field has a zero amplitude. This is consistent with our simulation results. We can verify this argument from the Hersenberg equation (\ref{eq:heisenbery_a}) of $\hat a$. Assuming that the cavity field adiabatically follows the atomic dynamics, the cavity field operator and photon number operator can be expressed by the atomic coordinate operator

\begin{eqnarray}
\hat a &=& \frac{iU_{\rm t}\sin(K \hat x)}{i\left[ \Delta_{\rm c}-U_0 \sin^2(K \hat x) \right] - \kappa}, \\
\hat a^\dag \hat a &=& \frac{U_{\rm t}^2\sin^2(K \hat x)}{\left[ \Delta_{\rm c}-U_0 \sin^2(K \hat x) \right]^2 + \kappa^2}.
\end{eqnarray}
From the expressions above we can see that $\hat a$ and $\hat a^\dag \hat a$ are odd and even functions of $\hat x$, respectively. Therefore, the mean value of the cavity field operator $\left\langle \hat a \right\rangle$, which is given by its integration over a period of the atomic spatial wave function, is zero for a symmetric initial state, while the average photon number $\left\langle \hat a^\dag \hat a \right\rangle$ is not. In an adiabatical limit the cavity field should be composed by coherent state components, and thus we can write these components as $|\alpha\rangle$, $|-\alpha\rangle$ and $|0\rangle$~\cite{oc273, epjd46}.

On the other hand, the atom can be recoiled with opposite phases by scattering a photon of the state $|\alpha\rangle$ or $|-\alpha\rangle$. The scattered field in the cavity provides an effective potential for the atom, which is composed by two parts: one is the dipole trap described by the $\sin^2(K\hat x)$ term (as shown by the sinusoidal curve in Fig.~\ref{fig:pn_coh_evol}a), and the other is due to the scattering force described by the $\sin(K\hat x)$ term. The field component with amplitude $\alpha$ ($-\alpha$) deepens the potential for, for example, the odd (even) sites, and raises the barriers for the even (odd) ones. Therefore, the atomic probability density forms two spatial modes, which are referred to as the odd mode and the even mode. We can define the correlation function as $\chi(x)=\int {\rm d}( K\xi ) |\rho_{\rm at}(\xi,\xi+x)|$ to describe the atomic spatial correlation property with distance $x$, with $\rho_{\rm at}(x_1,x_2)$ the reduced density matrix describing the atomic motion~\cite{jpb38}. The atomic spatial coherence can be obtained experimentally by detecting the cavity field intensity through the photons decay out of the cavity~\cite{prl102, oe18}. From Fig.~\ref{fig:pn_coh_evol}b we can see that, for the steady state, the coherence between neighboring sites is destroyed($\chi(\pi) \ll 1$), while the coherence between next-nearest-neighbor sites preserved. Therefore, the atomic density operator can be written as $\rho_{\rm at}= |{\rm odd}\rangle \langle{\rm odd}| + |{\rm even}\rangle \langle{\rm even}|$.

\begin{figure}[tbp]
\begin{center}
\includegraphics[width=7cm]{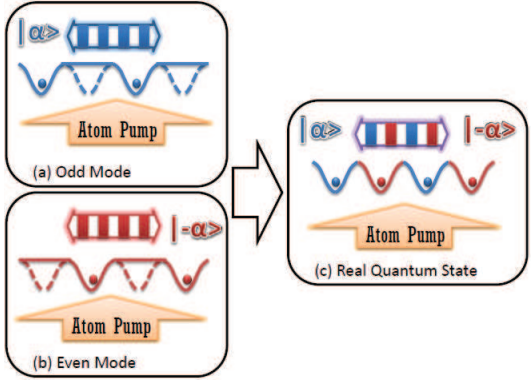}
\end{center}
\caption{(Color online) Schematic diagram of the atomic odd (a) and even modes (b), which are generated by scattering photons of the state component $| \alpha \rangle$ and $| -\alpha \rangle$, respectively. (c) shows the real quantum state composed of both the odd and the even modes. These two separated atomic modes scatter photons from pumping laser into the cavity and form two coherent state components with opposite phases.}
\label{fig:atom_modes_scheme}
\end{figure}

According to the analysis above, we can see that the scattering process changes the period of the atomic spatial distribution from $\lambda/2$ to $\lambda$, with $\lambda=2\pi/K$ the wavelength of the cavity field, and the atomic spatial coherence between neighboring sites is destroyed. This means that, two independent atomic spatial modes, coupling with two opposite coherent state components of the cavity field, respectively, are formed when the system gets into the steady state, which is shown in Fig.~\ref{fig:atom_modes_scheme}. The peaks of atomic spatial distribution centered at the odd sites form a matter-wave grating, which is referred to as the atomic odd spatial mode below (see Fig.~\ref{fig:atom_modes_scheme}a), and a coherent state component with amplitude $\alpha$ is coupled to the odd spatial mode. The peaks centered at even sites form the atomic even spatial mode, (see Fig.~\ref{fig:atom_modes_scheme}b), which is coupled to a coherent state component with amplitude $-\alpha$. The density operator of the system can be described as
\begin{eqnarray}\label{eq:two_modes_rho}
&\rho_{\rm fit}& = \epsilon \rho_{\rm at}^{({\rm O})} \otimes |\alpha\rangle \langle\alpha| + \epsilon \rho_{\rm at}^{({\rm E})} \otimes |-\alpha\rangle \langle-\alpha|\nonumber\\
&&+ (1-2\epsilon) \rho_{\rm at}^{({\rm R})} \otimes |0\rangle \langle0|.
\end{eqnarray}
Here $\rho_{\rm at}^{({\rm O})}$ ($\rho_{\rm at}^{({\rm E})}$) describes the atomic density matrix of the odd (even) mode and $|\alpha\rangle \langle\alpha|$ ($| \alpha\rangle \langle-\alpha|$) the field component scattered by the odd (even) mode. $\rho_{\rm at}^{({\rm R})}$ is the residual mode, coupled to the vacuum field $|0\rangle \langle0|$. This residual mode comes from the quantum property of the atom. Since the cavity field is weak, the atom is not well localized at any site, and there is still probability to find the atom at nodes of the cavity field. $\epsilon$ is the weight factor for the odd and even modes.

\subsection{Decomposition of atomic spatial modes and cavity field states}

We use the model described by Eq.~(\ref{eq:two_modes_rho}) to fit the steady-state density matrix $\rho_{{\rm ss}}$ from the Monte-Carlo simulation. First, we need to determine the amplitude of the two coherent state components $\alpha$ (or $-\alpha$) and the weight factor $\epsilon$, which can be calculated simply by accessing the information of the cavity field. The reduced density operator for the cavity field is given by $\rho_{{\rm cav}}(n_1, n_2)=\left\langle n_1 \right| \left( \sum_k \langle k | \rho_{{\rm ss}} | k \rangle \right) \left| n_2 \right\rangle$, and from Eq.~(\ref{eq:two_modes_rho}) we easily get
\begin{eqnarray}\label{eq:two_modes_rho_ph}
\rho_{{\rm cav\_fit}} = \epsilon \left| \alpha \right\rangle \left\langle \alpha \right| + \epsilon \left| -\alpha \right\rangle \left\langle -\alpha \right| + \left( 1-2\epsilon \right) \left| 0 \right\rangle \left\langle 0 \right|.
\end{eqnarray}
By using Eq.~(\ref{eq:two_modes_rho_ph}) to fit $\rho_{{\rm cav}}$, we have $\left\langle \hat a^2 \right\rangle = 2 \epsilon \alpha^2, \left\langle \hat a^4 \right\rangle = 2 \epsilon \alpha^4, \left\langle \hat a^\dag \hat a \right\rangle = 2 \epsilon \left| \alpha \right|^2$. Thus, $\pm\alpha$ are equal to the square roots of $\left\langle \hat a^4 \right\rangle / \left\langle \hat a^2 \right\rangle$, and the weight factor $\epsilon = \left\langle \hat a^\dag \hat a \right\rangle / \left( 2 \left| \alpha \right|^2 \right)$. The fitting error can be measured by ${{\rm Tr}}(\left| \rho_{{\rm cav\_fit}} - \rho_{{\rm cav}} \right|)$. Fig.~\ref{fig:MCWF_fit}a - \ref{fig:MCWF_fit}b shows the fitting results for the cavity photon statistics. With the increasing of the pumping strength, the average photon number in the cavity grows, and the fitting error becomes larger, which is due to the limited cutoff of the Fock basis for the cavity field. Nevertheless, as we can see in the figures, the fitted results (crosses) coincide well with the simulation results (bars) for different $U_{\rm t}$.

\begin{figure}[tbp]
\begin{center}
   \begin{picture}(0,0)
    \put(-10,80){(a)}\quad
     \end{picture}
   \includegraphics[width=3.9cm]{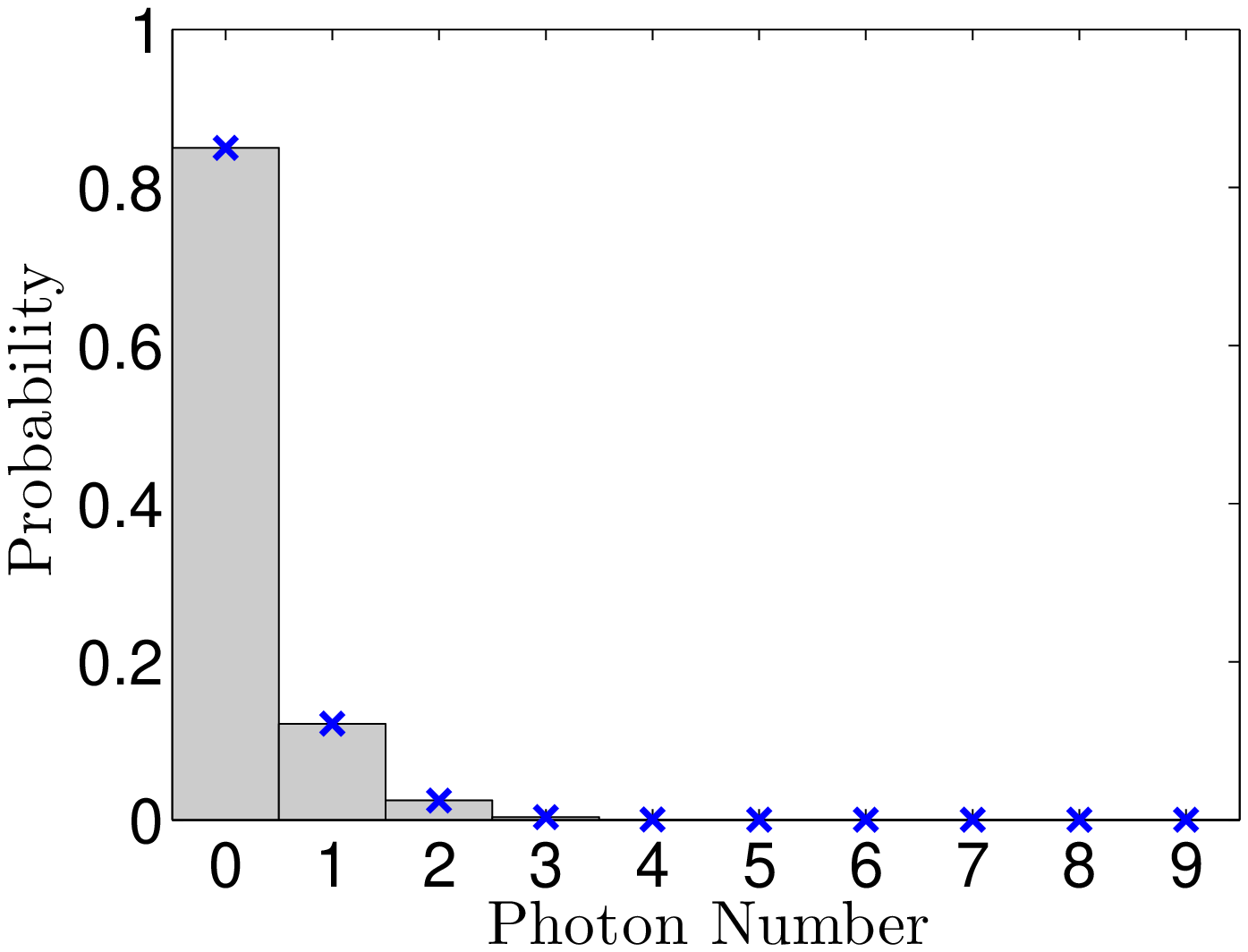}\quad\quad
   \begin{picture}(0,0)
    \put(-10,80){(c)}\quad
     \end{picture}
   \includegraphics[width=3.9cm]{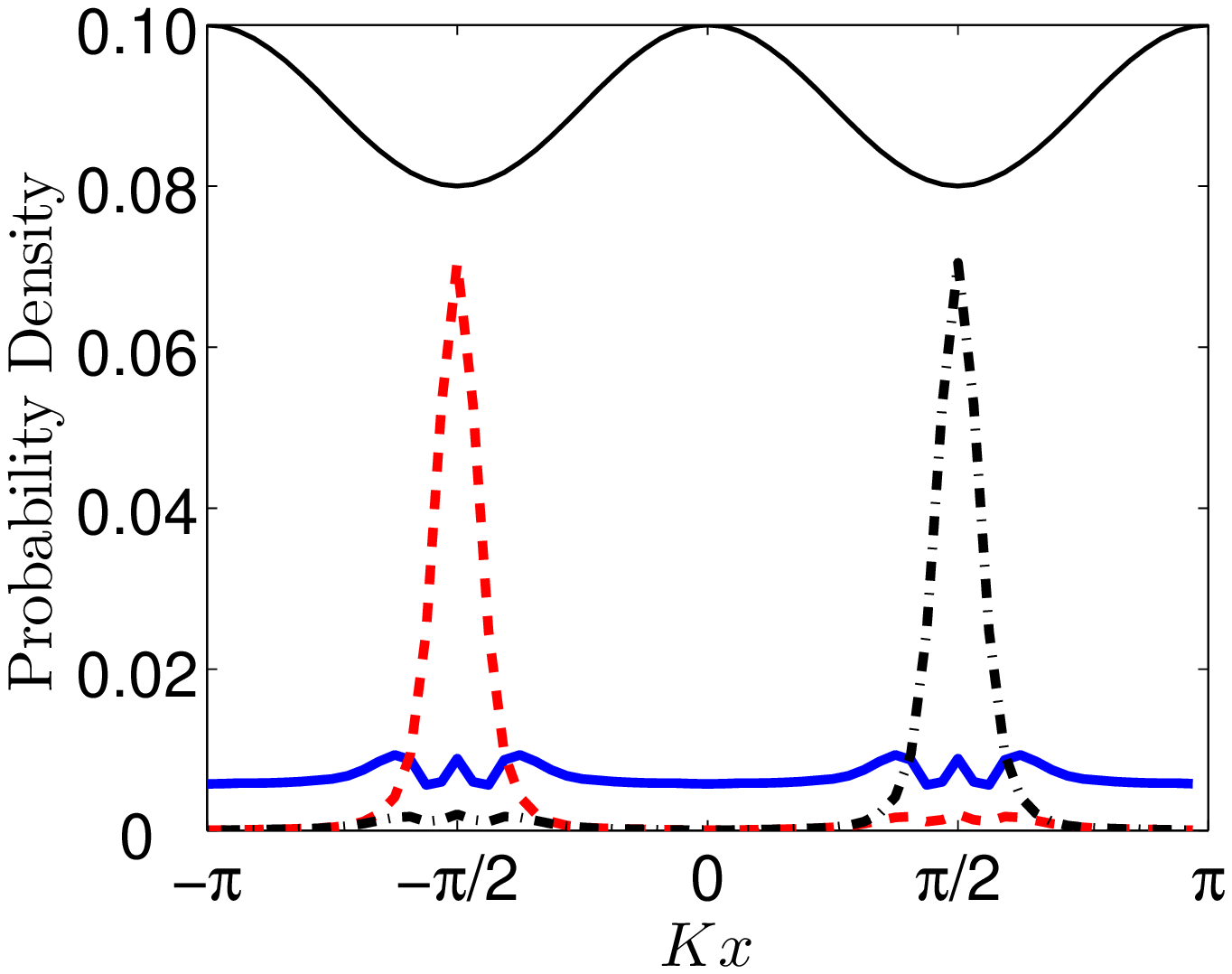}\quad\quad\\
   \begin{picture}(0,0)
    \put(-10,80){(b)}\quad
     \end{picture}
   \includegraphics[width=3.9cm]{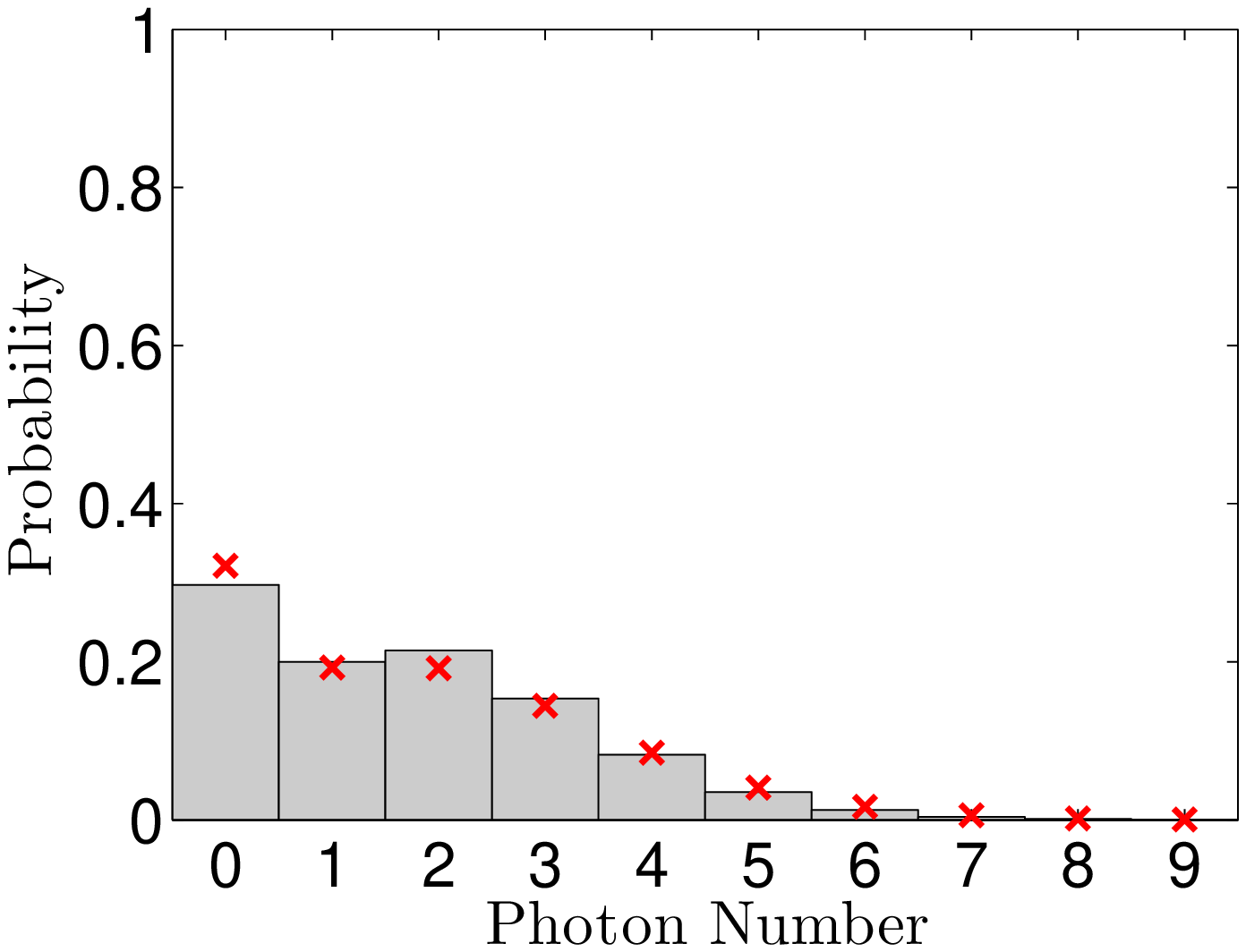}\quad\quad
   \begin{picture}(0,0)
    \put(-10,80){(d)}\quad
     \end{picture}
   \includegraphics[width=3.9cm]{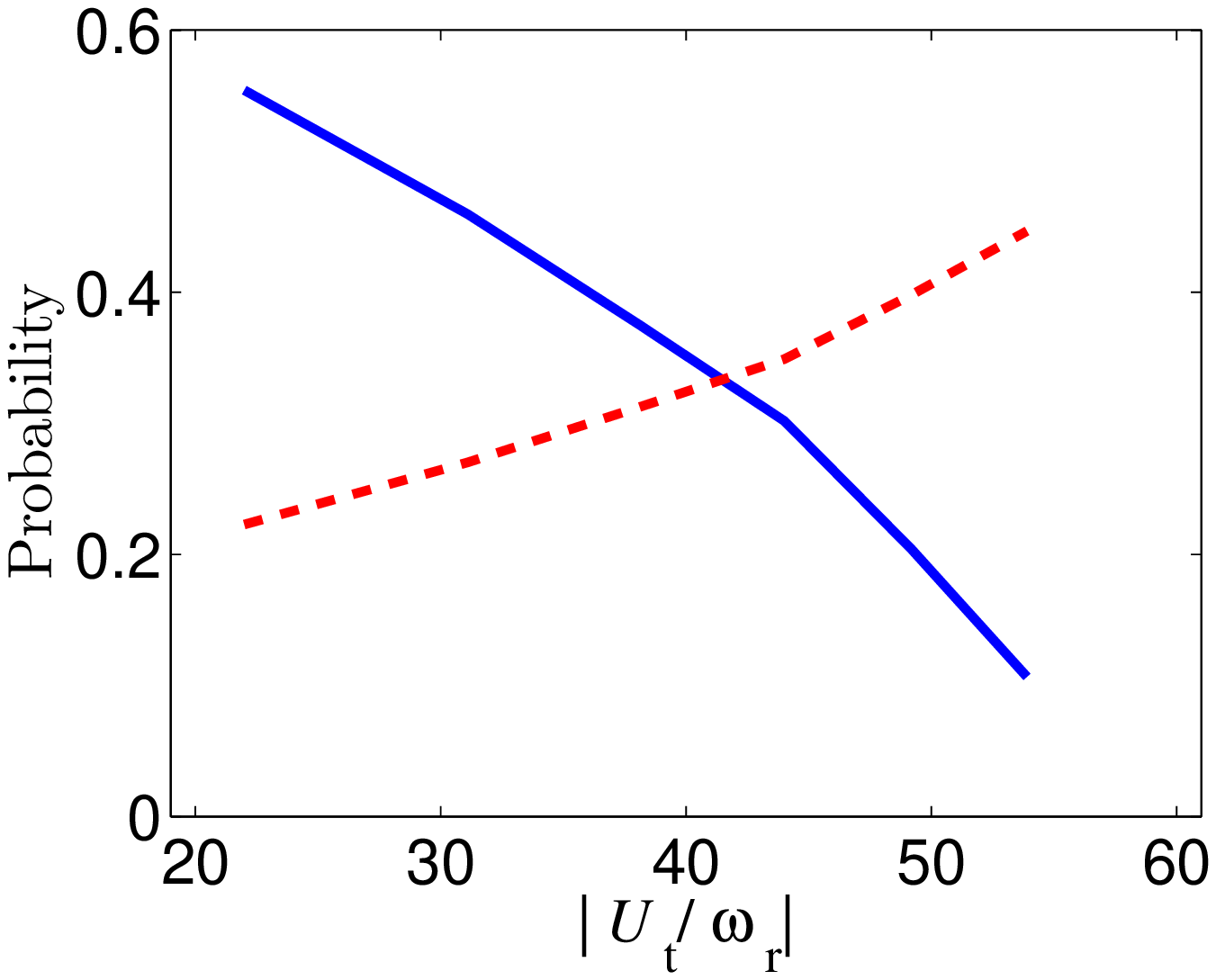}\quad\quad
\end{center}
\caption{(Color online) (a) and (b) show the photon number distribution from the Monte-Carlo simulation (bars) and the fit of the model described by Eq.~(\ref{eq:two_modes_rho_ph}) (crosses). The effective pumping strength $U_{\rm t}$ equals to $-22\omega_{\rm r}$ for (a) and $-49\omega_{\rm r}$ for (b). The error measured by ${\rm Tr}(\left| \rho_{{\rm cav\_fit}}-\rho_{{\rm cav}} \right|)$ is $0.13\%$ for (a) and $7.78\%$ for (b). (c) Fitted odd mode $\rho_{\rm at}^{({\rm O})}$ (dashed line), even mode $\rho_{\rm at}^{({\rm E})}$ (dash-dotted line) and residual mode $\rho_{\rm at}^{({\rm R})}$ (solid line) of the atomic spatial distribution for steady state with $U_t=-38\omega_r$. The solid sinusoidal curve shows the dipole trap potential in the cavity. (d) The probability of the residual mode (blue solid line) and the odd or even mode (red dashed line) for different $U_{\rm t}$. With the increasing of the pumping strength, the probability for the odd and even modes increases while that for the residual mode decreases. Other parameters are: $\kappa=31.25\omega_r$, $\Delta_c = U_0 = -390\omega_r$.}
\label{fig:MCWF_fit}
\end{figure}

By applying the operators $\hat a$ and $\hat a^2$ on both sides of Eq.~(\ref{eq:two_modes_rho}), and assuming that Eq.~(\ref{eq:two_modes_rho}) exactly describes the real density operator of the system $\rho_{{\rm ss}}$, the three parts in Eq.~(\ref{eq:two_modes_rho}) can be derived to be
\begin{eqnarray}
\epsilon \rho_{{\rm at}}^{({\rm O})} \left| \alpha \right\rangle \left\langle \alpha \right| &=& \frac{1}{2 \alpha^2} \left[ \hat a^2 \rho_{{\rm ss}} + \alpha \hat a \rho_{{\rm ss}} \right], \\
\epsilon \rho_{{\rm at}}^{({\rm E})} \left| -\alpha \right\rangle \left\langle -\alpha \right| &=& \frac{1}{2 \alpha^2} \left[ \hat a^2 \rho_{{\rm ss}} - \alpha \hat a \rho_{{\rm ss}} \right], \\
(1-2\epsilon) \rho_{{\rm at}}^{({\rm R})}\left| 0 \right\rangle \left\langle 0 \right| &=& \rho_{{\rm ss}} - \frac{1}{\alpha^2}\hat a^2 \rho_{{\rm ss}}.
\end{eqnarray}
By calculating the reduced density matrix for the atomic motion, $\rho_{{\rm at}}^{({\rm O})}$, $\rho_{{\rm at}}^{({\rm E})}$ and $\rho_{{\rm at}}^{({\rm R})}$ can be got. The fitted steady state spatial distribution for different atomic modes is plotted in Fig.~\ref{fig:MCWF_fit}c. We can see clearly the spatial distribution of the three atomic spatial modes. The residual mode is approximately uniform except for fluctuations due to the treatment error. This error follows the fact that in our treatment the $|\pm \alpha\rangle$ states are not independent with the vacuum state and there can be a small dip in the residual mode's density at the position of the peaks of the odd and even modes. The small peaks in the center of the dips of the residual mode may imply the existence of some higher-order modes. This residual mode implies that for a quantum particle, there is still certain probability to find it at the nodes. The weight factor $\epsilon$ describes how much of the atomic probability density has entered the ordered modes. Fig.~\ref{fig:MCWF_fit}d shows the dependence of the weights of the residual mode (blue solid line) and the ordered mode (red dashed line) on the pumping strength for the steady state results. We can see that with the increasing of the pumping strength, the probability for the residual mode becomes smaller, while that for the odd and even mode gets larger.

\section{Dynamics of the cavity field and atomic spatial modes}

\subsection{Dynamics of entanglement and cavity field properties}

In many-body physics an important goal is to reveal the connection between the entanglement and the quantum phase transitions. In previous work Ritsch et al has shown that the atom-field entanglement plays a crucial role during the process of self-organization. Here we investigate the entanglement of our fully quantum mechanical model and compare it with their results from a Bose-Hubbard-type model~\cite{oc273}. For convenience, we use a non-interacting BEC model to describe the physics. When illuminated by a transverse laser, the atoms scatter photons into the cavity. The state vector of the system immediately evolves into a superposition of two coherent states with opposite phases entangled to two ordered atomic states, respectively, that is, $| \psi \rangle = | {\rm odd} \rangle |\alpha\rangle + | {\rm even} \rangle |-\alpha\rangle$~\cite{oc273}. The entanglement can be measured by the ${{\rm negativity}}=\left|\sum_i \lambda_i \right|, \lambda_i<0$ with $\lambda_i$ the negative eigenvalue of the partial transposed density operator $\left\langle n_1,k_1 \right| \rho_{{\rm PT}} \left| n_2,k_2 \right\rangle \equiv \left\langle n_1,k_2 \right| \rho \left| n_2,k_1 \right\rangle$. Due to the cavity loss, the entanglement decays fast and the density matrix evolves into a mixed one as described by Eq.~(\ref{eq:two_modes_rho}), as shown in Fig.~\ref{fig:Q_and_neg}a. The evolution of the cavity field can be described by the statistical distribution of the photons, which can be reflected by the Mandel $Q$ factor defined as $Q=\left(\Delta\left( \hat a^\dag \hat a \right)-\left\langle \hat a^\dag \hat a \right\rangle\right)/\left\langle \hat a^\dag \hat a \right\rangle$. From Fig.~\ref{fig:Q_and_neg}b we can see that the $Q$ factor follows the evolution of the negativity, revealing the mutual influence between the atomic distribution and the cavity field. Finally, a stable optical lattice is established in the cavity consisting of two opposite coherent state components, and two ordered spatial modes of the atoms are formed, as shown in Fig.~\ref{fig:atom_modes_scheme}. The time evolution of negativity shows coincidence with the results from a Bose-Hubbard model~\cite{oc273}.

\begin{figure}[tbp]
\begin{center}
   \begin{picture}(0,0)
    \put(-10,80){(a)}\quad
     \end{picture}
   \includegraphics[width=3.9cm]{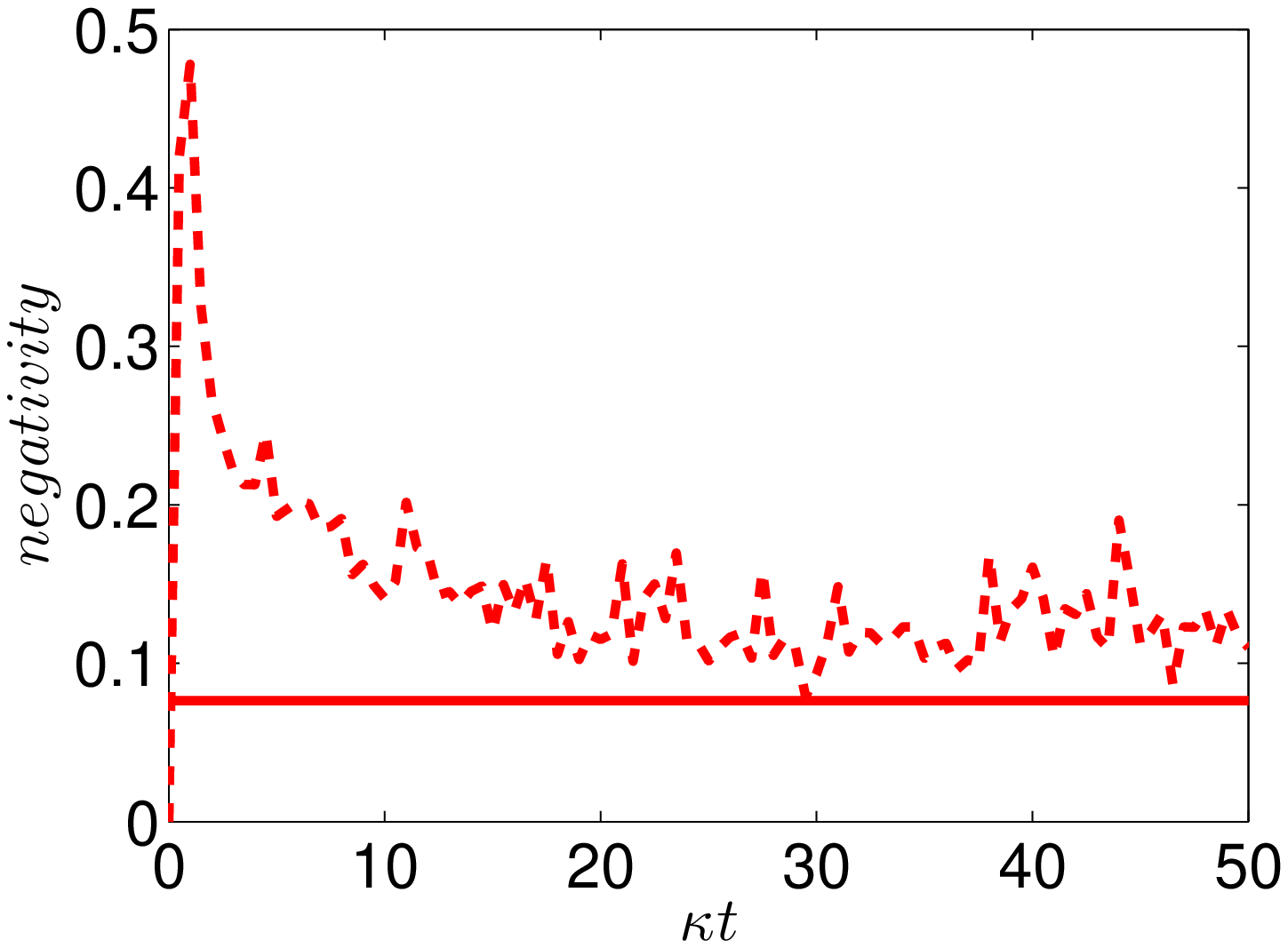}\quad\quad
   \begin{picture}(0,0)
    \put(-10,80){(b)}\quad
     \end{picture}
   \includegraphics[width=3.9cm]{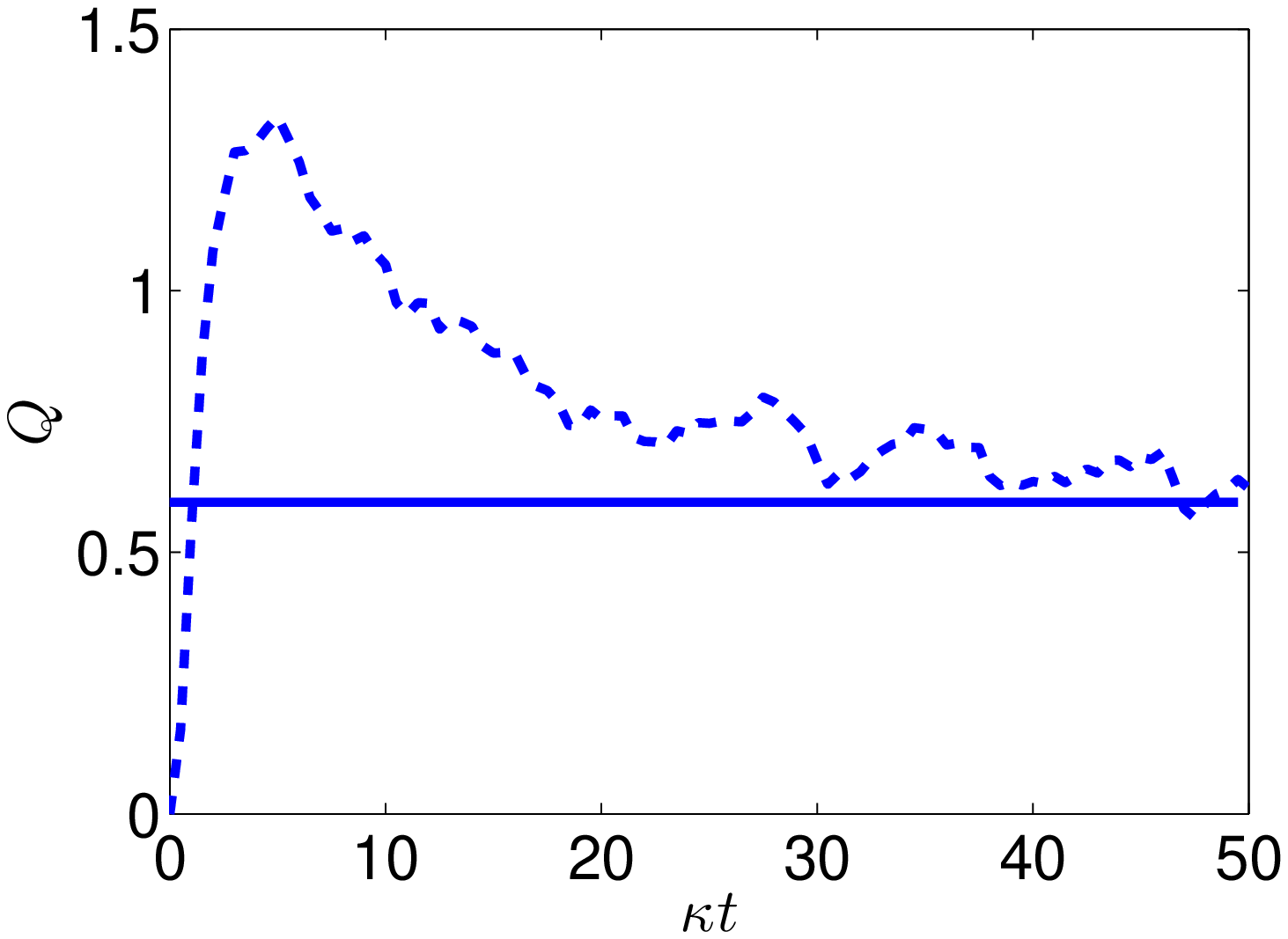}\quad\quad
\end{center}
\caption{(Color online) Time evolution (dashed line) and steady-state value (solid line) of negativity (a) and Mandel $Q$ factor (b) of the system. Parameters: $\kappa=31.25\omega_r$, $\Delta_c = U_0 = -390\omega_r$ and $U_{\rm t}=-49\omega_{\rm r}$.}
\label{fig:Q_and_neg}
\end{figure}

\subsection{Dynamical process for the formation of atomic ordered modes}

From Fig.~\ref{fig:Q_and_neg}a we can see that, since about $\kappa t=2$ the negativity decays smoothly and tends to its steady state value. As the cavity field adiabatically follows the atomic motion with our parameters, it is possible to use the method described in the previous section to deal for the steady state to deal with the time-dependent density matrix, assuming that the cavity field adiabatically follows the atomic dynamics. Fig.~\ref{fig:mode_dynam} shows the time evolution of the three atomic spatial modes. We can see clearly that the atomic odd and even modes are established gradually by consuming the residual mode.

\begin{figure}[tbp]
\begin{center}
   \begin{picture}(0,0)
    \put(-10,80){(a)}\quad
     \end{picture}
   \includegraphics[width=3.9cm]{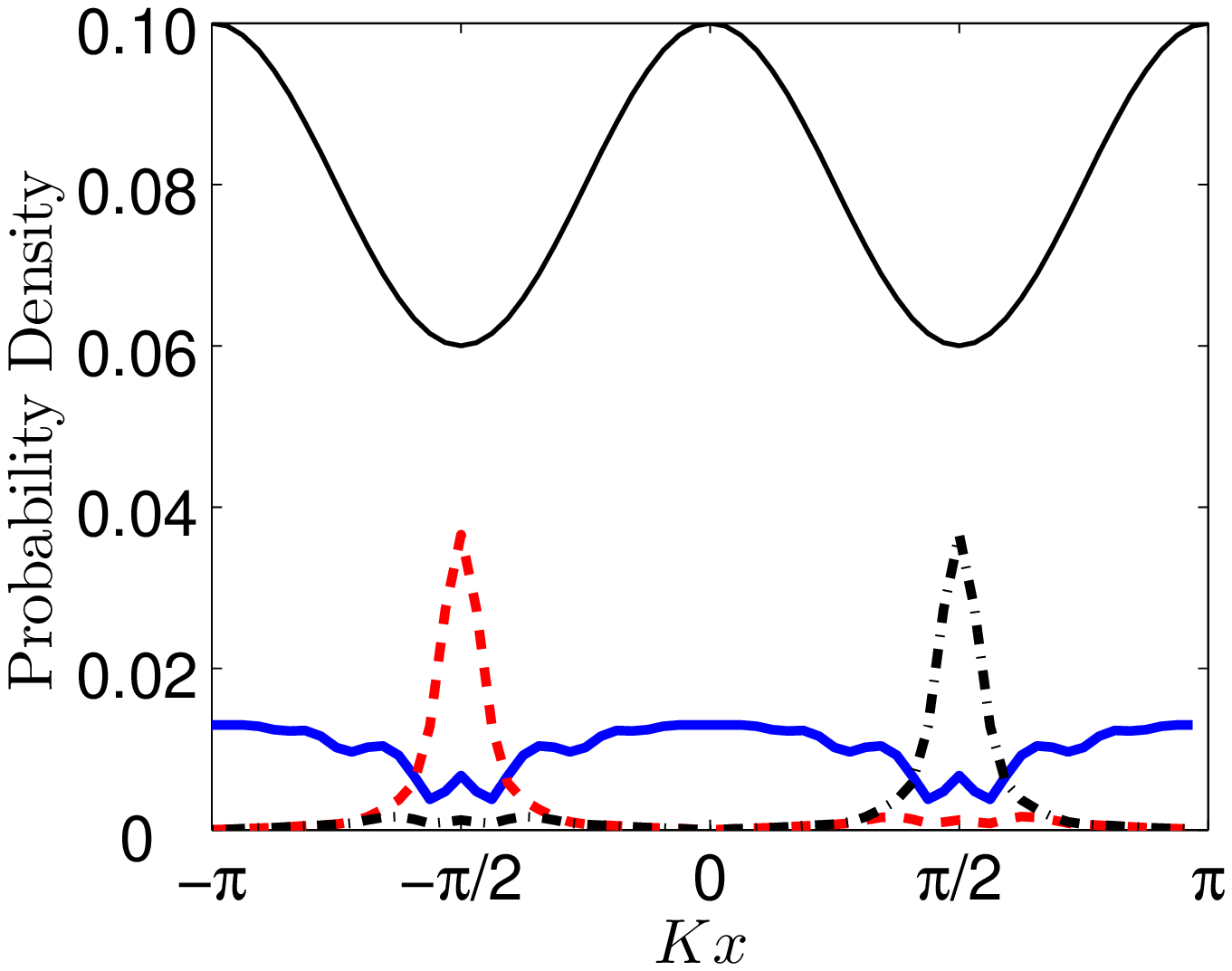}\quad\quad
   \begin{picture}(0,0)
    \put(-10,80){(b)}\quad
     \end{picture}
   \includegraphics[width=3.9cm]{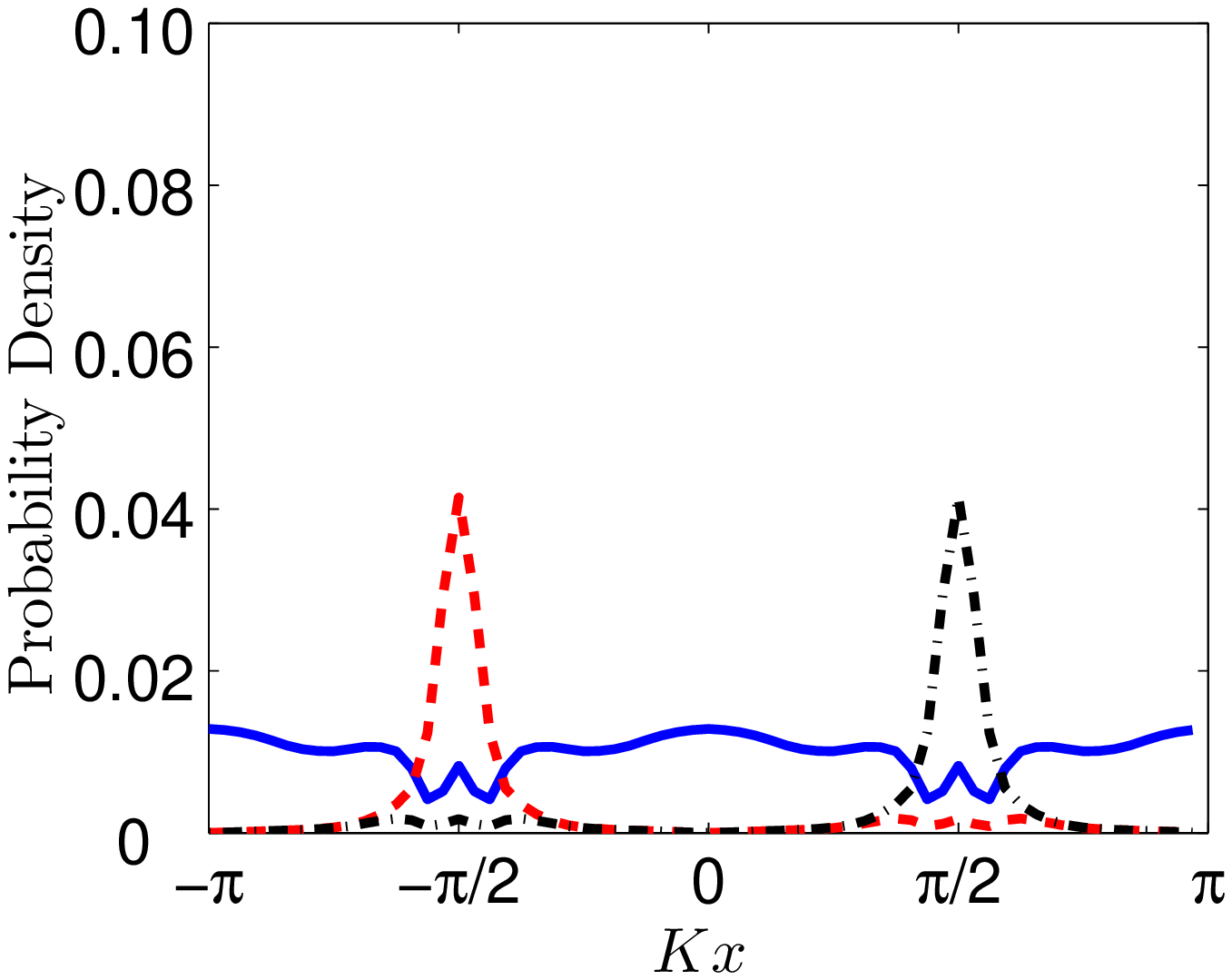}\quad\quad\\
   \begin{picture}(0,0)
    \put(-10,80){(c)}\quad
     \end{picture}
   \includegraphics[width=3.9cm]{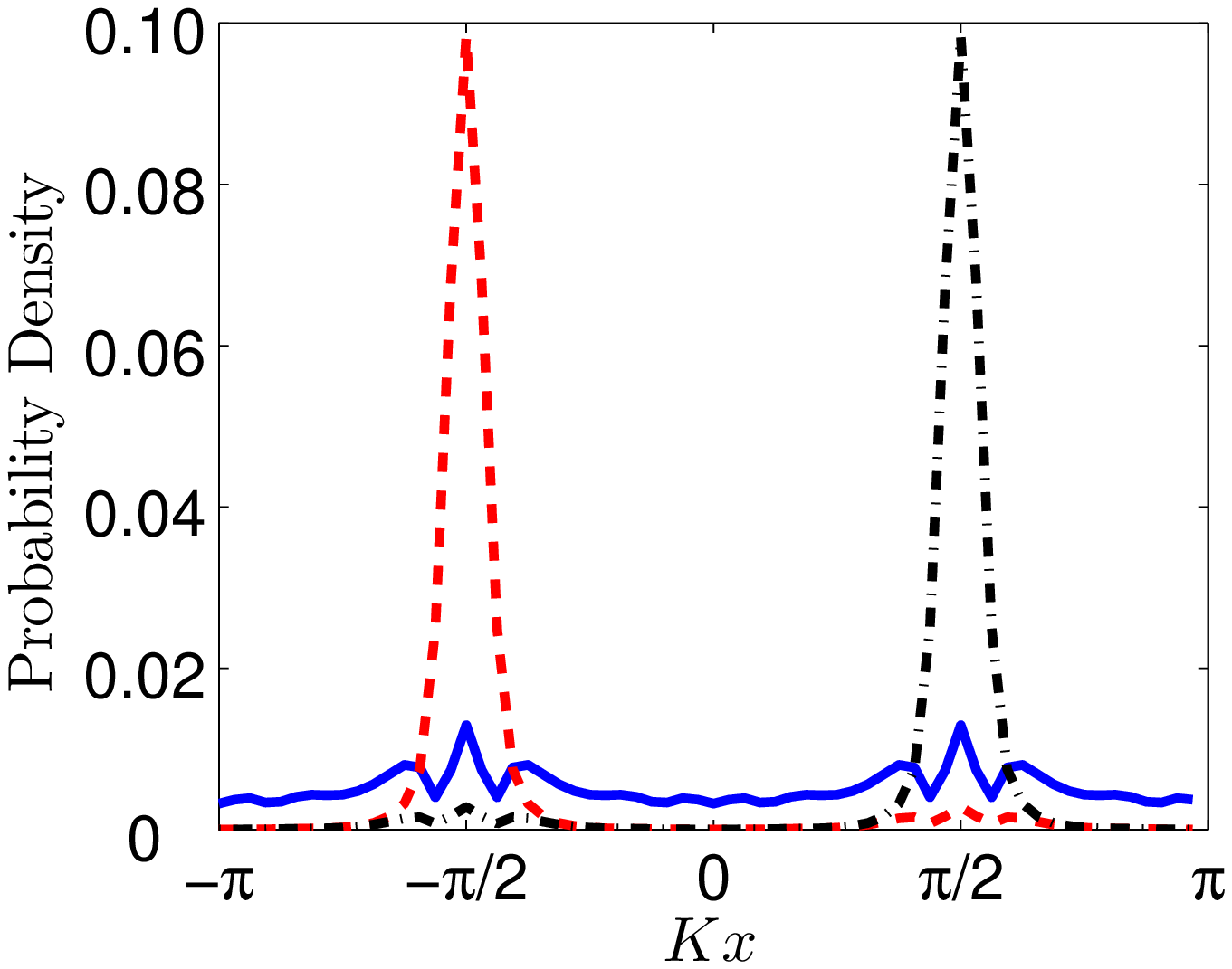}\quad\quad
   \begin{picture}(0,0)
    \put(-10,80){(d)}\quad
     \end{picture}
   \includegraphics[width=3.9cm]{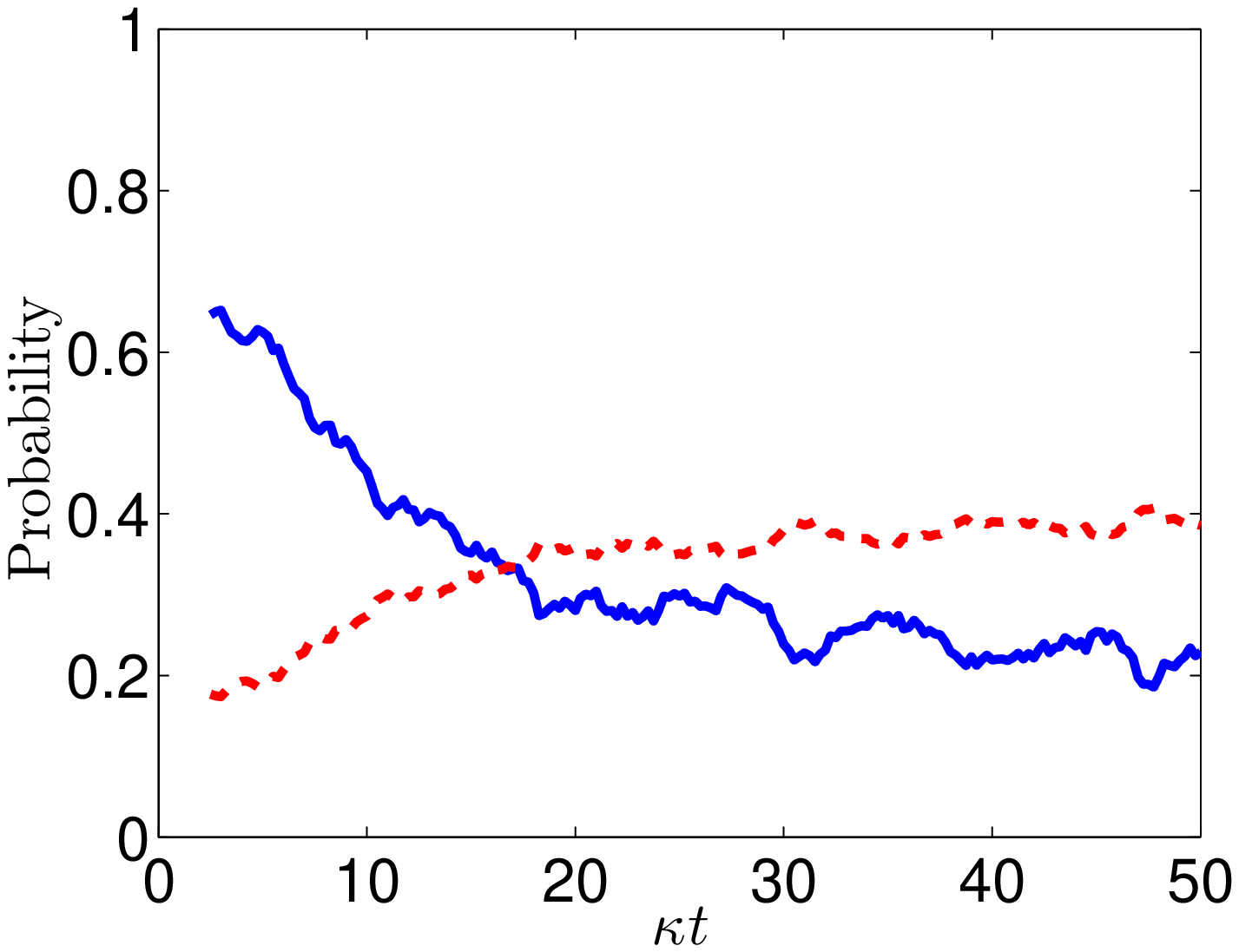}\quad\quad
\end{center}
\caption{(Color online) Time evolution of different atom modes. The figures (a)-(c) corresponds to the time $\kappa t$=2.5, 5 and 50. The red dashed line corresponds to the odd mode $\rho_{\rm O}^{\rm (a)}$, the black dot-dashed line to even mode $\rho_{\rm E}^{\rm (a)}$, and the blue solid line the resident distribution $\rho_{\rm R}^{\rm (a)}$. Parameters: $\kappa=31.25\omega_{\rm r}$, $\Delta_{\rm c}=U_0=-390\omega_{\rm r}$, $U_t=-49\omega_{\rm r}$. The solid sinusoidal curve in figure (a) shows the potential. (d) The time evolution of the probability of the homogeneous component (blue solid line) and the odd or even mode component (red dashed line) for $U_{\rm t}=-49\omega_{\rm r}$. We can see clearly that the odd and even mode arise from the uniform distribution.}
\label{fig:mode_dynam}
\end{figure}

\section{Conclusion}

In conclusion, considering the dissipation due to the cavity loss, we present a mixed state density operator model for the steady state of the system, which describes the atomic odd, even and residual spatial modes with cavity field states of amplitudes $\alpha$, $-\alpha$ and $0$, respectively. We use a decomposition treatment to get the spatial distribution of the atomic modes individually. According to the estimation of fitting error between the density operator and the simulation results, we find that this density operator is a good description of the system. By implementing this decomposition on the time-dependent simulation results, we show the dynamical process for the formation of the odd and even modes by consuming the residual mode. The mixed state density operator can be extended into a two dimensional model, and the atomic ordered spatial modes become the well-known ``checkerboard patterns'', which can be directly compared with experimental results. Besides, the model and the decomposition treatment can also be applied to the Bose-Hubbard-type model with an additional classical optical lattice potential in the Mott insulator regime, and the atomic modes may be quite different due to the presence of interatomic interaction. With quantum fluctuations and cavity field amplitude collapses, the atomic ordered modes can be self-organized to almost one of them, whose spatial or momentum distribution can be observed by in-situ imaging or time-of-flight (TOF) imaging technique. By inspecting the photons which decay out of the cavity, we can detect the field amplitude through homodyne detection, or the photon statistics with a single-photon-counting module (SPCM).

\section*{Acknowledgments}
Xiaoji Zhou thanks to J. Chen, H. W. Xiong and Z. W. Zhou for their help and discussion. We are also grateful to T. Vogt for thoroughly reading our manuscript. This work received support from the National Fundamental Research Program of China under Grant No. 2011CB921501, NSFC under Grant No. 61027016, No. 61078026 and No. 10934010.

\end{document}